# Galileo's Double Star: The Experiment That "Proved" the Earth Did Not Move

Christopher M. Graney
Jefferson Community College
1000 Community College Drive
Louisville, Kentucky 40272
(502) 213-7292
christopher.graney@kctcs.edu
www.jefferson.kctcs.edu/faculty/graney

ABSTRACT: Great opportunities arise for teaching physics, astronomy, and their histories when new discoveries are made that involve concepts accessible to students at every level. Such an opportunity currently exists thanks to the fact that notes written by Galileo indicating that he observed the double star Mizar in the "Big Dipper" have recently come to light. His measurements of this star, given the scientific knowledge at the time, strongly supported the theory that the Earth was fixed in space and not moving. Had Galileo published these results, it is likely that widespread acceptance of the heliocentric theory in scientific circles would have been significantly delayed. In light of these notes, his later reference in his *Dialogue* to using double stars as a means of proving that the Earth was in motion is puzzling. The physics and mathematics behind Galileo's work is easily within reach of students in introductory physics and astronomy courses, so discussion of Galileo's Mizar work and its interesting implications can be used in virtually any class.

# I. Introduction -- Galileo's Observations of Mizar

Great opportunities arise for teaching physics, astronomy, and their histories when new discoveries are made that involve concepts accessible to students at every level. Such an opportunity currently exists thanks to recent work by Leos Ondra that unearthed evidence that Galileo was among the first astronomers to observe a double star.[1] The first observation of a double star had been commonly attributed to the Jesuit astronomer Giambattista Riccioli in the middle of the 17th century. Ondra, suspicious that this was simply old lore passed from textbook to textbook, sought original sources to accurately date the first double-star observation. He came upon an obscure article by one Umberto Fedele, published in 1949 in a now-defunct journal, that indicated that Galileo -- together with Benedetto Castelli, a Benedictine mathematician and student of Galileo's -- observed the double star Mizar in the constellation of Ursa Major. Fedele's work led Ondra to notes written by Galileo that showed that Galileo had indeed observed Mizar, resolved its components, and measured their angular separation and sizes. Thus the first scientist known to observe the heavens with a telescope was also the first to observe a double star. However, Ondra inadvertently raises interesting and significant questions regarding Galileo's assumptions about the universe and the conclusions he drew from his observations -- questions that are accessible even to students in general "physics for poets" or "great ideas in science" classes, or in general astronomy classes.

    Ondra discusses how Galileo had a general interest in close groupings of stars; Galileo and Castelli studied the Trapezium in the constellation Orion as well as Mizar.

---

[1] Ondra, Leos, "A New View of Mizar," *Sky and Telescope*, July 2004, 72-75. An extended version of the *Sky and Telescope* article, with detailed references can be found at http://leo.astronomy.cz/mizar/article.htm.

While Galileo and Castelli communicated in vague terms, Ondra argues that Galileo hoped to use such groupings to find direct evidence that the Earth moved. Galileo supported the theory of Nicholas Copernicus that held that the Earth was moving around the Sun, but he lacked direct evidence of Earth's motion. Close groupings of stars were a logical place to look for the hard evidence he needed to convince skeptics who held that the Earth was stationary. Galileo certainly knew this. In 1611, not long after he began use of the telescope, he received a letter from one Lodovico Ramponi that described the use of a double star as an experimental test of whether the Earth moved and that encouraged Galileo to attempt just such a test.[2]

## II. Models/Theories in Play in Galileo's Time

Since the Copernican model is what we now know to be more correct, few physics and astronomy students (or professors, for that matter) understand just how firmly the skeptics in Galileo's time stood on solid scientific ground. During Galileo's time the view of the universe that had dominated Western thought for centuries was the view that the Sun, Moon, stars, and planets circled a stationary Earth. This "geocentric" theory, rooted in ideas expressed by Aristotle and refined by Ptolemy, postulated that heavenly bodies were made of an ethereal substance not found on Earth, and that their motion was powered by a "Prime Mover" that lay beyond the stars. Thus while on Earth it might be difficult just to keep a wagon moving along a road, the heavenly bodies could circle indefinitely because they were altogether different from anything found on Earth. It is difficult today for teachers or students to think like scientists half a millennium ago. At

---

[2] Siebert, Harald, "The Early Search for Stellar Parallax: Galileo, Castelli, and Ramponi," *Journal for the History of Astronomy*, Vol. 36 (2005) 251-271.

R1.1    3

that time the word "planet" did not mean a world such as Earth or an alien world like Tatooine in the movie *Star Wars*. The word "planet" meant an untouchable light in the heavens that wandered among the stars, that was made from an alien substance, and that had nothing in common with our Earth. Today, most physics or astronomy professors, if they discuss Aristotle at all, do so mainly to point out an example of bad physics, but a strong argument can be made that Aristotle's physics was much more compelling for its time than modern physicists realize and than Galileo admitted.[3]

      Galileo had rejected the geocentric view in favor of that of Copernicus, which postulated that the Earth was in fact a planet and that all the planets circled the Sun. This "heliocentric" theory offered some key advantages over the geocentric theory of Ptolemy. The Ptolemaic theory had explained the retrograde motions of the planets in terms of epicycles -- secondary circular motions superimposed upon the primary circular motions of planets around the Earth. The Ptolemaic theory could not provide a physical reason for why retrograde motion always occurred when a planet was in a line with the Earth and Sun. The mechanics of the Ptolemaic system could allow retrograde motion to occur at any point in a planet's motion, and the fact that it occurred when it did was simply a coincidence of nature. Copernicus' theory elegantly explained, via the planets having only a primary circular motion (around the Sun), both why and where retrograde motion occurred.[4]

---

[3] Barry Casper, "Galileo and the Fall of Aristotle: A Case of Historical Injustice?" *American Journal of Physics*, Vol. 45 (1977) 325-330. Casper argues that our view of Aristotle's physics has been distorted by problems in translation and changes in language.

[4] Douglas R. Martin, "Status of the Copernican Theory Before Kepler, Galileo and Newton," *American Journal of Physics*, Vol. 52 (1984) 325-330.

However, during Galileo's time there were major scientific problems with the Copernican theory. If the Earth did move, changes in perspective in our view of the stars due to our moving "observing platform" should manifest themselves over the course of a year. This effect, known as parallax, had not been detected. Furthermore, there was no known mechanism that could explain how a massive Earth made of very non-ethereal rock could move. The modern ideas in physics that explain how a massive world can move unceasingly when a wagon can't easily be kept moving down a road had not been developed in Galileo's time. Thus Tycho Brahe had proposed a hybrid theory in which the Sun, Moon, and stars circled a fixed Earth while the planets circled the Sun. Brahe's geocentric theory was mathematically identical to the Copernican theory and thus retained the Copernican theory's advantages.[5] But it left key advantages of the Aristotelian theory intact: no expectation of seeing parallax; no need to explain how the massive Earth could move; and ethereal heavens powered by a Prime Mover.

## III. The Lack of Conclusive Data

With his telescope Galileo observed the phases of Venus. The phases and Venus's changing size as seen through the telescope were completely consistent with Venus circling the Sun, and completely inconsistent with the Ptolemaic theory. These observations, announced in December of 1610, may have demolished the Ptolemaic theory, but they were fully compatible with Brahe's geocentric theory as well as with Copernicus's heliocentric theory. In both of those theories Venus circled the Sun. The phases of Venus, and the rest of Galileo's observations, illustrated problems with

---

[5] Johannes Kepler (Aiton, Duncan, & Field, translators), *The Harmony of the World* (American Philosophical Society/Diane Publishing Company, Darby, Pennsylvania,1997) pp. 403-404.

Aristotle and Ptolemy's ideas, but did not contradict Brahe's ideas. They were not direct evidence that the Earth actually moves.[6]

This lack of direct evidence made it tough to persuade intelligent skeptics. In a letter of April 15, 1615, to one Paolo Foscarini, the Roman Catholic Cardinal Robert Bellarmine wrote

> … if there were a true demonstration that the sun is at the center of the world and the earth in the third heaven, and that the sun does not circle the earth but the earth circles the sun, then one would have to proceed with great care in explaining the Scriptures that appear contrary, and say rather that we do not understand them than that what is demonstrated is false. But I will not believe that there is such a demonstration, until it is shown me….

Bellarmine goes on to write that being able to explain the motions of the planets by supposing that the Earth circles the Sun is not the same as direct evidence that the Earth actually moves.[7] Bellarmine may have been skeptical, but his letter to Foscarini had given Galileo an opening; if Galileo could find direct experimental evidence of a moving Earth, Bellarmine would be open to changing his mind. So too might other skeptics.

## IV. Galileo's Mizar Observations -- Conclusive Data

Based on what Galileo knew at the time, a close grouping of stars such as Mizar or the Trapezium should have provided the direct experimental evidence he needed. A group of stars at differing distances, closely spaced, would reveal the parallax and thus provide the direct evidence that would convince skeptics that Copernicus was right. Ondra's article dates Galileo's notes on Mizar and the Trapezium to early 1617 -- shortly after Bellarmine's letter to Foscarini.

---

[6] An easily available reference for students interested in learning more about Galileo's work is Rice University's Galileo Project (http://galileo.rice.edu/).
[7] The full text of Bellarmine's letter to Foscarini is available at http://www1.bellarmine.edu/strobert/about/foscarini.asp.

Students with only a basic knowledge of geometry can easily follow Galileo's notes and their ramifications. Galileo's notes indicate the apparent size of Mizar's two component stars to be six and four arc-seconds, with a gap between them of ten arc-seconds (corresponding to a center-to-center separation of fifteen arc-seconds). Galileo's observations were good, agreeing with modern measurements for the separation of the two component stars. Just how good an observer Galileo was has been illustrated by amateur astronomers Tom Pope and Jim Mosher. Pope and Mosher constructed a Galilean telescope and took digital photographs through the telescope.[8] Comparing their work to Galileo's highlights Galileo's skill and dispels any thoughts students might have that a 17th-century physicist's work would be crude [Figure 1].

Galileo's notes show that he used a very logical method to determine the distances to Mizar A and B. He first assumed the two stars were roughly the same size as the Sun. Using geometry, he then determined that since Mizar A's apparent size of six arc-seconds was 1/300 of the Sun's apparent size of 1800 arc-seconds (half a degree), Mizar A must be 300 times more distant than the Sun (in modern parlance, 300 A.U.).[9] The distance to the smaller Mizar B would then be 450 A.U.

This geometric method works perfectly in determining distances to church steeples, people, ships, etc. It was implicitly used in arguing that the size of Venus (along with its phase) showed that the planet circled the Sun. Galileo could not have known that stars were the one thing for which his geometric method of measuring distance would not work.  Galileo, being the first to ever use a telescope to observe the stars, knew nothing of light waves from a point source diffracting through a circular aperture.  In fact, the size of the Airy disk formed by diffraction through a telescope with the aperture of his is a few arc-seconds.  Pope and Mosher have reproduced Galileo's measurements of the sizes and separations of Mizar A and B both with their telescope and by doing a computer simulation for a telescope of size similar to one known to be used by Galileo [Figure 2].  They show that the sizes Galileo measured had everything to do with waves, optics, diffraction, and Airy disks and nothing to do with the actual sizes of the two stars.  But Galileo could not have known this.  He had every reason to believe that his size measurements were good, and that his distance calculations were as good as his assumption that the stars were suns.

---

[8] "CCD Images From a Galilean Telescope" http://www.pacifier.com/~tpope.
[9] Ondra, *S&T* (ref. 1), pp. 74-75.

Galileo must have expected Mizar to provide conclusive proof that Earth was in motion.  Based on his calculations of the distance to Mizar A, he would have expected it to have a parallax angle of about ±11.5 arc-minutes, with Mizar B having a parallax angle of ±7.6 arc-minutes.[10]  Thus their motion relative to each other would be on the order of several arc-minutes -- dwarfing the fifteen arc-second separation of the two components.  Galileo must have thought he would see the components of Mizar swing around each other dramatically as he observed them over a period of weeks and months [Figure 3].

But in fact Mizar A and B do not budge.  There was no way for Galileo to know that the physics of light waves hid the true size of Mizar A and B; no way for him to know that they were impossibly far away; no way for him to know that therefore the parallax effect he was looking for was impossibly small (so small that it would not be

---

[10] Seibert, *JHA* (ref. 2), p.269.  See endnote 78.

R1.1                                                8

detected in any star for another two hundred years). Galileo's precise measurements and logical calculations predicted that Mizar A and B would swing around each other dramatically over the course of a few months, providing conclusive evidence that the Earth was moving. Galileo must have been incredibly disappointed when subsequent observations of Mizar showed no change in positions at all. Since no parallax could be observed in Mizar, Galileo logically had to conclude that either the Earth was not moving, or that his assumption that the stars were suns at differing distances from Earth was wrong by orders of magnitude.

## V. Double Stars and the *Dialogue*

Yet Galileo asserts both that the Earth is moving and that the stars are suns at differing distances from Earth when, a decade and a half after observing Mizar and the Trapezium, he published his *Dialogue Concerning the Two Chief World Systems*. In the "Third Day" of the *Dialogue* Galileo argues (through the character of Salviati) that first magnitude stars have an apparent size of 5 arc-seconds (or 300 "thirds," where a third is one sixtieth of an arc-second), and that sixth magnitude stars have an apparent size of 50 thirds or 5/6 arc-seconds -- sizes in a line with his Mizar observations. He then assumes that stars are the same size as the Sun. And finally, based on that assumption, he proceeds to argue that, since 5/6 arc-seconds is 1/2160 the size of the Sun, sixth-magnitude stars are 2160 A.U. distant. This follows the calculations he used with Mizar.[11]

---

[11] Galileo Galilei, *Dialogue Concerning the Two Chief World Systems – Ptolemaic & Copernican (2nd edition)*, translated by Stillman Drake (University of California Press, Los Angeles, California, 1967), pp. 359-360.

Through the character of Salviati, Galileo also compares the yearly displacement of the stars caused by Earth's motion (i.e. parallax) to the retrograde motion of the outer planets. He discusses how the retrograde motion of outer planets is similar to but larger than what can be expected to be seen among the stars. He points out that these motions of the planets are noticeable by way of comparison to the background of stars, but that the parallax of the stars will not be easily seen because there is not a further background for them to be compared against.[12] He then goes on to say

> …it is not entirely impossible for something some time to become observable among the fixed stars by which it might be discovered what the annual motion does reside in. Then they, too, no less than the planets and the sun itself, would appear in court to give witness to such motion in favor of the earth. For I do not believe that the stars are spread over a spherical surface at equal distances from one center; *I suppose their distances from us vary so much that some are two or three times as remote as others*. Thus *if some tiny star were found by the telescope quite close to some of the larger ones, and if that one were therefore very remote, it might happen that some sensible alterations would take place among* them corresponding to those of the outer planets [italics added]….[13]

This reads as though Galileo never observed Mizar at all. Mizar A and B already had appeared "in court" in the manner he was proposing. To the best of his knowledge they had already given witness -- that the Earth did <u>not</u> move and that the "annual motion" resided in the Sun.

## VI. Questions for Discussion

Since the view Galileo promotes in the *Dialogue* is in direct conflict with his earlier work on Mizar that Ondra has brought to light, many interesting questions arise that can provide fodder for classroom discussion. Apparently Galileo found the Copernican model -- the Earth circling the Sun which in turn was one of many suns -- so compelling that he stood by it in the face of strong experimental evidence against it, even when that evidence was his own observations.[14] What made the Copernican model more compelling to Galileo than his own data will make excellent material for class discussion.

---

[12] Galileo, *Dialogue* (ref. 11), pp. 381-382.
[13] Galileo, *Dialogue* (ref. 11), pp. 382-383.

Another excellent avenue for discussion is that while Galileo may have sat on his Mizar observations, today a scientist might be expected to publish such data. Suppose Galileo had, in fact, published his results. What would have been the effect of that publication on the ongoing scientific debate regarding the Earth's motion? Brahe's geocentric theory was already more compatible with the physics of the early 17th century than was the Copernican theory. It remained a force in the scientific debate even after Galileo's death. If Galileo's Mizar observations had come to light, it seems likely that it would have taken longer for the heliocentric theory to be widely accepted; certainly some scholars, unlike Galileo, would have found the Mizar data more compelling than the Copernican model. After all, direct evidence of the motion of the Earth did not come until more than a century after Galileo observed Mizar and the Trapezium, when James Bradley came upon the phenomenon of Stellar Aberration in the 1720's. By that time the fundamentals of modern "Newtonian" physics existed and explained how the massive Earth could move.

A third interesting line of discussion might be what Galileo's opponents would have thought had they known that, at the time he published the *Dialogue*, Galileo was sitting on results that strongly challenged the Copernican theory he was so adamantly championing? Since Galileo shared work with many in the Roman Catholic Church (such as Castelli), did perhaps some of Galileo's adversaries know of his Mizar work, or at least suspect as much, and thus sharpened their knives all the more quickly when the *Dialogue* was published?

---

[14] Siebert finds that Galileo and Castelli communicated about at least four different close star groupings over a span of a decade. These included, besides Mizar and the Trapezium (which they observed in early 1617), a grouping between Orion and Canis Major which Castelli observed in early 1617, and one in Scorpius which Castelli found in 1627 (Siebert, *JHA* (ref. 2), pp. 260-262).

      Ondra's work settled the question of who observed the first double star, but it opened many more questions. Happily the concepts behind these questions are accessible even to students in the most introductory courses. The topic of Ondra's discoveries and what they mean can surely enrich many a "physics for poets" class -- and will probably make for interesting discussion even in more advanced classes and seminars.

**Figure 1**

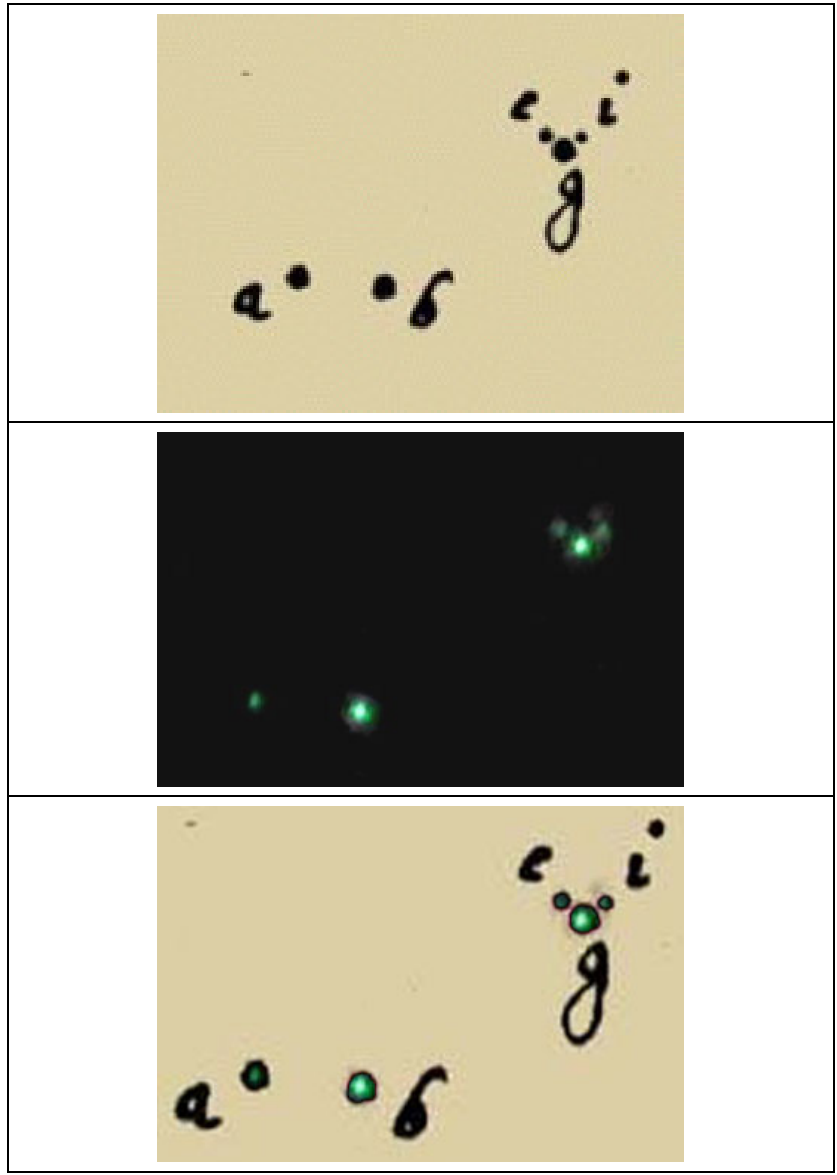

Figure 1 caption:
**Top to Bottom: Galileo's drawing of the Trapezium and nearby stars; Pope and Mosher's image of the same region obtained using a Galilean telescope and digital camera; Pope and Mosher's superposition of Galileo's drawing over their image, illustrating the accuracy with which Galileo recorded his observations. Images used with permission. This figure may appear in color online.**

**Figure 2**

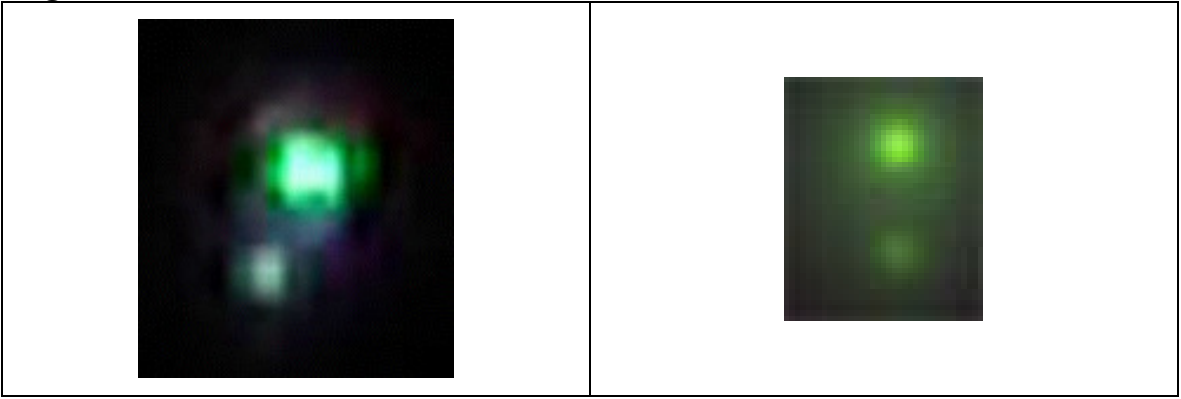

Figure 2 caption:
**Pope and Mosher's photo (left) and ray-trace simulation (right) showing Mizar A and B as Galileo's telescopes would have shown them. The resolution is a little better in the simulation as it corresponds to a 38 mm telescope actually used by Galileo, whereas the photo is taken through a slightly smaller telescope. Images used with permission. This figure may appear in color online.**

**Figure 3**

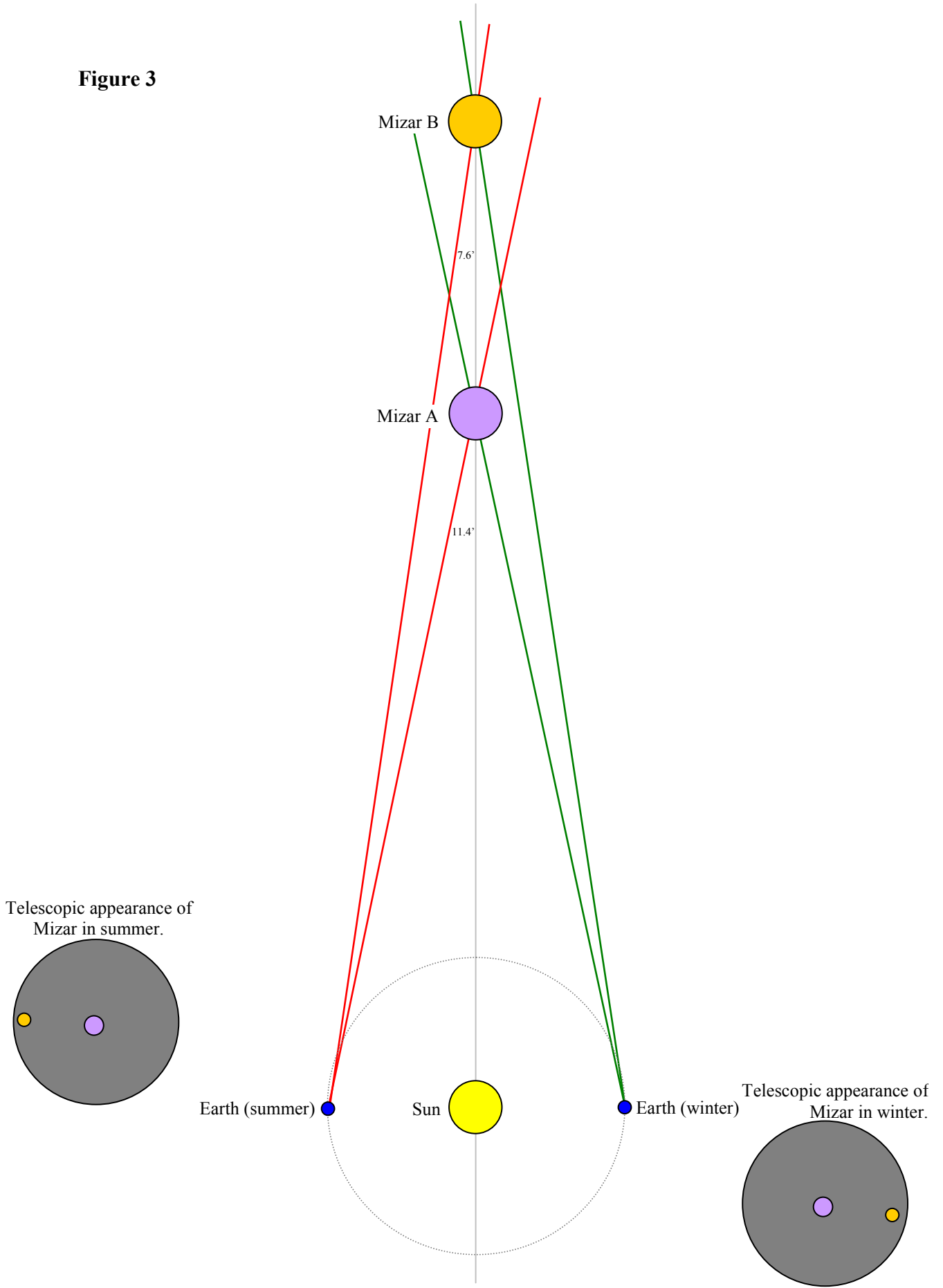

Figure 3 caption:

**Earth, the Sun, and the two components of Mizar according to Galileo's measurements and assumptions. Mizar A and B are both the size of the Sun. The Sun-Mizar A distance is 300 times the Earth-Sun distance. The Sun-Mizar B distance is 450 times the Earth-Sun distance. The difference in the parallax angles of the two stars is such that the relative positions of the two stars should alter dramatically as the Earth swings from one side of its orbit to the other – by far more than the separation between the components observed by Galileo – and should have been easily detectable. The change in the stars' positions would have been hard evidence that the Earth was in motion. Galileo did not know that the sizes of star images in his telescopes were a function of wave optics and not of geometry, so his distance estimates to Mizar A and B were far too small. This figure may appear in color online.**